\documentstyle{pss}
\input epsf.tex
\begin{document}

\title{On the symmetry of a Preisach map}

\author{{\sc Marek W. Gutowski}\footnote{) Corresponding author;
e-mail: gutow@ifpan.edu.pl} $^{)}$ }

\address{Institute  of  Physics,   Polish   Academy   of   Sciences, 
 Al.~Lotnik\'ow 32/46,\\ 02--668 Warszawa, Poland}

\submitted{\today
} \maketitle

\hspace{9mm} Subject classification:
07.05.Tp; 
74.25.Ha; 
75.10.-b; 
75.40.Mg; 
75.60.Ej; 

\begin{abstract}
    At the very  heart  of  the  successful  phenomenological  model  of 
 magnetic hysteresis there is the so called  Preisach  distribution.  In 
 the  existing  literature  it  is   implicitly   assumed,   that   this 
 distribution has a mirror symmetry. We show, by simple  and  convincing 
 example, that this common assumption is plainly wrong. Dropping it,  we 
 gain the ability to model not only the usual  hysteresis  loops  (major 
 and minor) more accurately than ever before, but also those  displaying 
 the exchange bias effect, what is impossible within  the  framework  of 
 the symmetrical Preisach model.  It  is  hoped,  that  our  observation 
 paves the  way  towards  the  unified  description  of  all  hysteretic 
 systems, including, but not necessarily  limited  to,  superconductors, 
 (multi)layered structures, nanocrystalline materials, patterned  media, 
 and -- perhaps -- the other non-magnetic hysteretic phenomena.
\end{abstract}

\section{Introduction}
    The major hysteresis  loop,  observed  in  the  sizable  samples  of 
 homogeneous ferromagnetic materials, exhibits well known symmetry:
\begin{equation}\label{obvious}
    M_{\rm lb}(H) = -M_{\rm ub}(-H),
\end{equation}
    where $M_{\rm lb}$ ($M_{\rm ub}$) denotes the lower  (upper)  branch 
 of the sample's magnetization, $M$\ {\em  vs.\/}  exciting  field  $H$. 
 The hysteresis curve, not necessarily the  major  loop,  but  also  the 
 response to the arbitrary sequence of exciting fields as well,  can  be 
 described, or modelled, in many ways. One of them, the  {\em  Classical 
 Preisach  Model\/}  (CPM),  was  first  proposed  by  Ferenc   Preisach 
 \cite{Ferenc} and then subsequently developed, generalized  and  tested 
 by many  researchers.  In  this  model,  the  change  of  the  sample's 
 magnetization is expressed by the double integral:
\begin{equation}\label{up-sweep}
    \Delta M = M(H_{\rm f}) - M(H_{\rm i}) = 2M_{\rm s}\!\!\!\!
  \mathop{
  \int\int}_{{H_{\uparrow}\ge H_{\downarrow}} \atop 
  {H_{\rm i} \le H_{\uparrow} \le H_{\rm f}}}\!\!\!\!dH_{\uparrow}\,
  dH_{\downarrow}\ \varrho\left( H_{\uparrow}, H_{\downarrow} \right),
\end{equation}
    for the monotonously increasing field (`i' -- the initial state, `f' 
 -- final) and by
\begin{equation}\label{down-sweep}
    \Delta M = M(H_{\rm f}) - M(H_{\rm i}) = -2M_{\rm s}\!\!\!\!
  \mathop{
  \int\int}_{{H_{\uparrow}\ge H_{\downarrow}} \atop 
  {H_{\rm f} \le H_{\downarrow} \le H_{\rm i}}}\!\!\!\!dH_{\uparrow}\,
  dH_{\downarrow}\ \varrho\left( H_{\uparrow}, H_{\downarrow} \right),
\end{equation}
    if the field is monotonously decreasing. $M_{s}$ is  the  saturation 
 magnetization,  while  the  distribution  $\varrho\left(  H_{\uparrow}, 
 H_{\downarrow}\right)\ge    0$,\    supported    over    the     domain 
 $H_{\uparrow}\ge H_{\downarrow}$ is called the Preisach  density.  With 
 the additional condition:
\begin{equation}
    \mathop{\int\int}_{}
    dH_{\uparrow}\,dH_{\downarrow}\ \varrho\left(H_{\uparrow},
  H_{\downarrow}\right) = 1,
\end{equation}
    the  Preisach  density  is  commonly   regarded   as   a~probability 
 distribution. It describes  the  probability  of  encountering  the  so 
 called {\em hysteron\/}, i.e. the hypothetical object characterized  by 
 elementary,  rectangularly  shaped,  hysteresis  loop  with   up-   and 
 down-switching fields equal  to  $H_{\uparrow}$  and  $H_{\downarrow}$, 
 respectively.


    It is obvious, that the symmetry  presented  in  Eq.~(\ref{obvious}) 
 will be preserved, if the Preisach distribution,  $\varrho$,\  is  also 
 symmetric, namely when
\begin{equation}
    \varrho\left( H_{\uparrow}, H_{\downarrow} \right)  =  \varrho\left( 
 -H_{\downarrow}, -H_{\uparrow}\right)
\end{equation}
    holds for any pair $\left( H_{\uparrow}, H_{\downarrow} \right)$  of 
 its arguments, $H_{\uparrow}\ge H_{\downarrow}$.  The  mirror  symmetry 
 of $\varrho\left( H_{\uparrow}, H_{\downarrow}  \right)$  with  respect 
 to the line $H_{\downarrow} = -H_{\uparrow}$ is therefore commonly {\em 
 assumed\/}, before any reconstruction of the Preisach  distribution  is 
 ever attempted. In this paper we are  going  to  show,  that  symmetric 
 hysteresis loops {\em can\/} be generated  by  Preisach  distributions, 
 which themselves have no mirror symmetry.
\begin{figure}[h]
\begin{tindent}
\epsfxsize=7cm
\epsffile{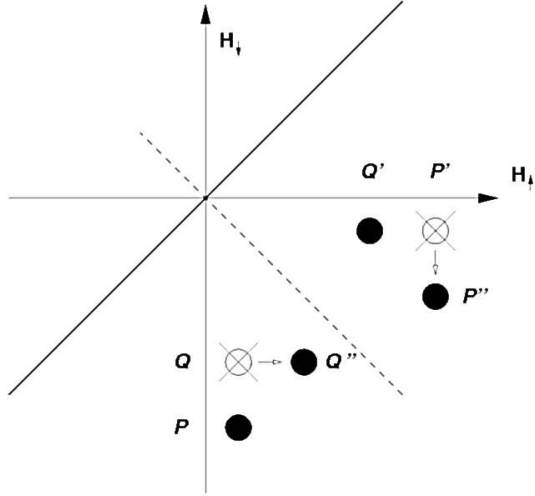}
\end{tindent}
\caption{
    An example  of  the  non-symmetric  Preisach  map,  which  generates 
 symmetric major hysteresis loop. Initially the only components  of  the 
 map are two  symmetrically  located  pairs  of  positive,  delta-shaped 
 peaks, with identical amplitudes,  labeled  as  $\left(P,Q\right)$  and 
 $\left(P^{\prime},Q^{\prime}\right)$ respectively
}
\label{fig1}
\end{figure}
    In other words:  while  the  mirror  symmetry  of  $\varrho$  indeed 
 implies the experimentally observed symmetry properties  of  hysteresis 
 loops, then the inverse  need  not  to  be  true.  The  idea  that  the 
 Preisach  distribution  has  to  be  symmetric  is  present   in   the 
 literature since almost $50$ years \cite{Biorci-Pescetti58}.  It  seems 
 so obvious that nobody ever cared to prove it formally.
For example, Mayergoyz (1986) wrote in \cite{May0}
\begin{quote}{\sl
    \ldots\     \     it     can     be     easily     proved     that}\ 
 $\mu\left(\alpha,\beta\right) = \mu\left(-\beta,-\alpha\right)$
\end{quote}
    however he didn't  present  the  proof  (Mayergoyz's  $\mu$  is  our 
 $\varrho$).

\section{Hints for possible asymmetry}
    Let us try to prove it now by reduction {\em ad absurdum\/}.  Assume 
 that we have a nonsymmetric  distribution  and  let's  see  whether  it 
 necessarily has to generate  the  non-symmetric  hysteresis  loop.  The 
 distribution is non-symmetric, if  there  exists  at  least  one  point 
 $\left( H_{\uparrow}^{\star}, H_{\downarrow}^{\star}  \right)$  in  its 
 support, such that the condition  $\varrho\left(  H_{\uparrow}^{\star}, 
 H_{\downarrow}^{\star}         \right)         =          \varrho\left( 
 -H_{\downarrow}^{\star}, -H_{\uparrow}^{\star} \right)$  is  violated.\ 
 But if this indeed is the case,  then  the  integrals  (\ref{up-sweep}) 
 and (\ref{down-sweep}) remain unchanged, since  a  single  point  is  a 
 zero-measure set on a Preisach plane!\ As we can see, it is  impossible 
 to prove the well known claim. One could argue that our finding has  no 
 physical meaning, since such kind of asymmetry cannot  be  detected  in 
 any real experiment and, consequently, it should be regarded only as  a 
 mathematical toy. Eventually we might  modify  the  original  claim  to 
 read: {\em the Preisach distribution is symmetric almost everywhere  on 
 a Preisach plane\/}. However, we are not going to stop  at  this  point 
 and shall present a~well-founded, physical counterexample.
\begin{figure}[h]
\begin{tindent}
\epsfxsize=7cm
\epsffile{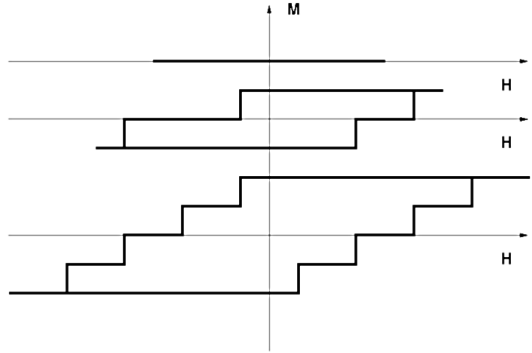}
\end{tindent}
\caption{
    Stacked  hysteresis  loops  generated  by   non-symmetric   Preisach 
 distribution presented in Fig.~\ref{fig1}  for  various  amplitudes  of 
 the exciting field $\left[-H,+H\right]$. For small amplitudes there  is 
 a null response (hence the loop  is  symmetric),  in  the  intermediate 
 regime the loop becomes asymmetric and exhibits the so called  exchange 
 bias effect, while the major loop (lowest) is symmetric again
}
\label{fig2}
\end{figure}
\section{Derivation of the main result}
    Our counterexample uses four Dirac's delta  functions  as  the  sole 
 components of a Preisach distribution. This  is  not  completely  crazy 
 idea, since the delta peaks are the only mathematical entities able  to 
 reproduce well known Barkhausen jumps.  Those  jumps  appear  sometimes 
 very stable against repeated magnetization reversals. The same is  true 
 for  the  domain   structures   \cite{memory,memory-bis}.   The   other 
 advantage is that the single-point support of a delta  function  is  no 
 longer  a  zero-mesure   set   for   integrals   (\ref{up-sweep})   and 
 (\ref{down-sweep}).\ We are not original  introducing  delta  functions 
 into  Preisach  model  (see  Pescetti  \cite{Pescetti}  1989),  so  our 
 derivation cannot be viewed only as a mathematical curiosity.

    Consider  Fig.~\ref{fig1}.  Initially  the   Preisach   distribution 
 consists of two delta peaks of  equal  amplitudes,  located  at  points 
 labeled as $P$ and $Q$, and their symmetrical images  $P^{\prime}$  and 
 $Q^{\prime}$, respectively. Such a distribution gives rise, of  course, 
 to symmetrical  hysteresis  loops.  But  now  we  destroy  the  initial 
 symmetry  by  moving  points   $Q\rightarrow\   Q^{\prime\prime}$   and 
 $P^{\prime}\rightarrow\  P^{\prime\prime}$,  as  indicated  by  arrows. 
 Having done  so,  we  can  observe  (Fig.~\ref{fig2})  that  the  major 
 hysteresis loop,  as  well  as  some  minor  loops,  remain  symmetric, 
 although their overall shape has changed. There  are  also  some  minor 
 loops, which are asymmetric now, imitating the so called exchange  bias 
 effect.

    It is also possible  to  rearrange  the  four  initial  delta  peaks 
 differently, in such a way as to preserve the symmetry of  {\em  all\/} 
 minor loops with turning  points  of  equal  magnitudes.  This  can  be 
 achieved    by    rotating    the    pairs    $\left(P,Q\right)$    and 
 $\left(P^{\prime},Q^{\prime}\right)$    around    points    $P$     and 
 $Q^{\prime}$ respectively by the same  angle  $\varphi$.  Let  initialy 
 $P=\left(H_{0},-H_{2}\right)$,   $Q=\left(H_{0},-H_{1}\right)$,    with 
 $d=\vert H_{1}-H_{2}\vert$. After rotation,  the  jumps  on  the  lower 
 branch   of   the   hysteresis   loop   ccur   at   fields:    $H_{0}$, 
 $H_{0}+d\sin\varphi$, $H_{1}$ and $H_{1}+d\cos\varphi$\ (in  increasing 
 order). Upper branch  jumps  (in  decreasing  order)  are  located  at: 
 $-H_{0}$, $-H_{0}-d\sin\varphi$, $-H_{2}+d\cos\varphi$  and  $-H_{2}$.\ 
 One can easily see, that in all cases  $H_{\rm  lb}^{\rm  jump}=-H_{\rm 
 ub}^{\rm jump}$, what proves that all the relevant symmetry  properties 
 of the hysteresis loops are indeed preserved.

\section{Comments}
    Our  construction  strongly  suggests  that  the  true  (asymmetric) 
 Preisach distribution may be a~conformal image of  some  `ideal',  i.e. 
 symmetric map. It is worth to note, however, that conformal mapping  of 
 the Preisach triangle onto itself, if  applicable  at  all,  cannot  be 
 arbitrary. Ordinary Euclidean transformations,  i.e.  translations  and 
 rotations,  are  useless  (shifted  distribution  fails   to   describe 
 exchange biased loops, see \cite{Ktena}).
    It  remains  unclear,  whether  the  asymmetry   of   the   Preisach 
 distribution alone would be able to  describe  correctly  the  recently 
 reported   \cite{Negative}   hysteresis   loops   exhibiting   negative 
 remanence.

\section{Conclusions}
    We have shown that, contrary to  the  common  belief,  the  Preisach 
 distribution need not  to  be  symmetric.  This  seems  to  remove  the 
 apparent  disagreement  between  the   distributions   recovered   from 
 experimental    data    on    one     of     the     classical     ways 
 \cite{Hoff88,Hoff89,Verdi} (which implicitly assume  the  existence  of 
 symmetry questioned  here,  sometimes  even  including  the  postulated 
 particular shape of the map to be reconstructed) with those obtained by 
 the  recently   introduced   FORC   (First   Order   Reversal   Curves) 
 \cite{Pike99,PikePRB} diagrams technique, which reveals  the  asymmetry 
 quite  often;   see   \cite{Katzgraber,Stancu2003,Katzgraber2004}   and 
 especially impressive figures presented by Robb,  Novotny  and  Rikvold 
 \cite{Robb}.

    The  construction  of  FORC  diagrams  implicitly   suppresses   the 
 reversible part of the Preisach map, i.e. the one located on  the  line 
 $H_{\uparrow}=H_{\downarrow}$. In our opinion, it is the combination of 
 both techniques, what should constitute the complete tool for physical, 
 as opposed to purely  phenomenological,  characterization  of  magnetic 
 systems and their internal interactions.

\end{document}